\documentclass[a4paper,11pt]{article}
\pdfoutput=1 

\usepackage{jheppub} 

\usepackage[T1]{fontenc} 
\usepackage[utf8]{inputenc}
\usepackage{amsmath,amssymb,amstext,amsfonts} 

\usepackage[dvipsnames]{xcolor}
\usepackage{graphicx}
\usepackage{hyperref}
\usepackage{cancel} 
\usepackage{float}
\usepackage{bbm}
\usepackage{mathtools}

\usepackage{lscape}
\usepackage{siunitx}
\usepackage{subcaption}

\usepackage{cleveref}

\usepackage{slashed}
\usepackage{physics}
\usepackage{nicematrix}

\makeatletter
\renewcommand*\env@matrix[1][\arraystretch]{%
  \edef\arraystretch{#1}%
  \hskip -\arraycolsep
  \let\@ifnextchar\new@ifnextchar
  \array{*\c@MaxMatrixCols c}}
\makeatother

\makeatletter
\g@addto@macro\bfseries{\boldmath}
\makeatother

\newcommand{\eps}{\epsilon}

\newcommand\scalemath[2]{\scalebox{#1}{\mbox{\ensuremath{\displaystyle #2}}}}

\def\beq{\begin{equation}}
\def\eeq{\end{equation}}
\def\bsp#1\esp{\begin{split}#1\end{split}}
\def\dim{{\mathbf D}^-}

\usepackage{tikz}
    \usetikzlibrary{positioning}
    \usetikzlibrary{arrows}
    \usetikzlibrary{shapes}
    \usepackage{pgfplots}
    \usetikzlibrary{calc}
    \usetikzlibrary{decorations.markings}
     \usetikzlibrary{decorations.pathmorphing}
    \tikzset{snake it/.style={decorate, decoration=snake}}
\def\centerarc[#1](#2)(#3:#4:#5) 
    { \draw[#1] ($(#2)+({#5*cos(#3)},{#5*sin(#3)})$) arc (#3:#4:#5); }
    \usepackage{booktabs}

\definecolor{cnblue}{RGB}{7,82,154}

\title{\boldmath 
On the photon self-energy to three loops in QED}

\author[a]{Felix Forner,} 
\author[a]{Christoph Nega,} 
\author[a]{and Lorenzo Tancredi} 

\affiliation[a]{Technical University of Munich, TUM School of Natural Sciences, Physics Department, James-Franck-Straße 1, D-85748 Garching, Germany}

\emailAdd{felix.forner@tum.de}
\emailAdd{c.nega@tum.de}
\emailAdd{lorenzo.tancredi@tum.de}

\preprint{\begin{minipage}[t]{8cm}\begin{flushright} 
TUM-HEP-1539/24
      \end{flushright}\end{minipage}}

\abstract{

We compute the photon self-energy to three loops in Quantum Electrodynamics. The method of differential equations for Feynman integrals and a complete $\epsilon$-factorization of the former allow us to obtain fully analytical results in terms of iterated integrals involving integration kernels related to a K3 geometry. We argue that our basis has the right properties to be a natural generalization of a canonical basis beyond the polylogarithmic case and we show that many of the kernels appearing in the differential equations, cancel out in the final result to finite order in $\epsilon$. We further provide generalized series expansions that cover the whole kinematic space so that our results for the self-energy may be easily evaluated numerically for all values of the momentum squared. From the local solution at $p^2=0$, we extract the photon wave function renormalization constant in the on-shell scheme to three loops and confirm its agreement with previously obtained results. 
}

\begin{document} 
\maketitle
\flushbottom


\section{Introduction}
\label{sec:intro}
Within any given Quantum Field Theory (QFT), two-point correlators are among the most fundamental quantities one can consider. From a practical point of view, in a renormalizable QFT, two-point Green’s functions provide some of the elementary ingredients required for the perturbative UV renormalization of the theory, see for example~\cite{Broadhurst:1991fi,Melnikov:2000zc}. These include building blocks for the calculation of the $\beta$-functions of the theory, which describe the running of the couplings with the energy scale, and the conversion factors required to study the renormalization-scheme dependence of higher-point correlators. From a more formal perspective, two-point correlators 
are the simplest correlators that exhibit a non-trivial dependence 
on external kinematics and, as such, can be used as a proxy for the lower bound of the mathematical complexity expected to appear at a given loop order in the perturbative expansion. Given the basic role they play, two-point correlators in different QFTs have received a lot of attention in the literature.

Quantum Electrodynamics (QED) is arguably the simplest of the physically relevant gauge theories, and it is no surprise that the photon and electron self-energies, which are the one-particle irreducible building blocks of QED two-point functions, were among the first calculations ever attempted to higher loop orders. In fact, their evaluation was pushed to the two-loop order already more than 60 years ago by Källen and Sabry~\cite{Kallen:1955fb} and Sabry~\cite{Sabry}, respectively, and the first few orders in the expansion close to zero momentum squared for the photon propagator to three loops where obtained in~\cite{Baikov:1995ui}. While the two-loop photon self-energy could already then easily be obtained in terms of classical polylogarithms, its electron counterpart famously provided the first example of a class of Feynman integrals in perturbative QFT that required the generalization of polylogarithms to an elliptic geometry. The origin of these new mathematical structures can be easily identified in the so-called two-loop massive sunrise graph~\cite{Bauberger:1994by, Bauberger:1994hx, Bauberger:1994nk, Caffo:1998du, 
Laporta:2004rb, Kniehl:2005bc, Remiddi:2016gno, Adams:2013nia, Adams:2014vja, Adams:2015gva, Adams:2016xah, Bloch:2013tra}, whose study in the past two decades has provided one of the main guides to generalize the theory of polylogarithms to the elliptic case~\cite{Levin:2007tto, Brown:2011wfj, Broedel:2014vla, Adams:2014vja, Passarino:2016zcd, Remiddi:2017har, Broedel:2017kkb, Broedel:2018qkq}.
Since then, elliptic geometries have been encountered in a variety of two-loop Feynman graphs, both in massive and massless theories~\cite{Aglietti:2007as, Sogaard:2014jla, vonManteuffel:2017hms, Bonciani:2016qxi, Adams:2018kez}, including in the highest supersymmetric cousin of 
Quantum Chromodynamics~\cite{Caron-Huot:2012awx,Bourjaily:2017bsb,McLeod:2023qdf}. 

Interestingly, it turns out that by increasing the number of loops 
or mass scales in the considered problems, generalizations of elliptic geometries become relevant, most notably their higher-dimensional Calabi-Yau (CY)~\cite{Bloch:2014qca, Bloch:2016izu, Vanhove:2014wqa, Primo:2017ipr, Bezuglov:2021jou, Kreimer:2022fxm, Klemm:2019dbm, Bonisch:2020qmm, Duhr:2022dxb, Bourjaily:2018ycu, Bourjaily:2018yfy, Bourjaily:2019hmc, Vergu:2020uur, Duhr:2022pch, Duhr:2023eld, Duhr:2024hjf} and higher-genus~\cite{Huang:2013kh, Marzucca:2023gto} counterparts. More recently similar integrals related to CY geometries have also been observed in the context of the two-body scattering problem in general relativity~\cite{Bern:2021dqo, Frellesvig:2023bbf, Klemm:2024wtd, Driesse:2024feo}. Once more, it turns out that the photon and electron self-energies in QED at three loops and beyond are among the simplest correlators where an entire class of these new geometries is expected to appear.
In fact, it has been known for some time that the generalizations of 
the sunrise graph to higher loops, i.e., the so-called $\ell$-loop banana integrals, can be expressed in $D=2-2\epsilon$ space-time dimensions in terms of iterated integrals over periods of an $(\ell-1)$-fold CY geometry~\cite{Bonisch:2021yfw}. 
Working in QED with one single massive lepton, which we will refer to as the electron for simplicity, it is then easy to convince oneself that, 
due to fermion-number conservation, at each new order in the 
perturbative expansion the very same $(\ell-1)$-fold Calabi-Yau manifolds can become relevant for the calculation of either the electron or the photon self-energy, depending on whether $\ell$ is even or odd. 
More explicitly, as we already discussed, the two-loop electron self-energy is the first to receive a contribution from the sunrise elliptic curve (a CY one-fold), while the photon self-energy remains polylogarithmic. Moving one loop higher, it was recently demonstrated~\cite {Duhr:2024bzt} that at three loops, the electron self-energy still depends on the same elliptic curve as its
lower-loop version. On the other hand, the three-loop photon self-energy 
admits the equal-mass banana graph as a subgraph and is, therefore, potentially the simplest two-point correlator in QED with an explicit dependence on the corresponding CY two-fold, a one-parameter K3 surface~\cite{Primo:2017ipr}.
We recall here that Feynman integrals with similar analytic properties have been considered previously in the literature. In particular, cuts of the same integrals appeared in the calculation of the QED photon spectral density \cite{Onishchenko:2022yvs} and in QED cross sections~\cite{Lee:2019wwn,Lee:2020mvt}, while very recently, cuts of similar graphs have been computed in the context of the evaluation of the $H \to b\bar{b}$ decay in QCD~\cite{Wang:2024ilc}.

At variance with the references above, in this paper, our aim is to calculate the full set of uncut integrals contributing to the QED self-energy to the three-loop order fully analytically as a function of the momentum squared. The main advances that make this calculation possible for us are the recent breakthroughs in the construction of $\epsilon$-factorized bases of differential equations for arbitrary geometries~\cite{Adams:2016xah,Adams:2018bsn,Pogel:2022ken,Pogel:2022vat,Giroux:2022wav,Frellesvig:2021hkr,Dlapa:2022wdu,Frellesvig:2023iwr,Gorges:2023zgv,Pogel:2022yat,Behring:2023rlq}.
In particular, we employ the algorithmic construction described in~\cite{Gorges:2023zgv}, which also provides us with a further non-trivial testing ground to demonstrate its general applicability to solve realistic problems in perturbative QFT. Importantly, our differential equations are not only in $\epsilon$-factorized form, but their matrix of differential forms satisfies two further restrictions. First of all, we make sure that \emph{locally} close to each regular singular point of the original equations, our differential equations admit at most single poles. Secondly, as part of our construction, we verify that all differential forms are linearly independent, which allows us to write solutions in terms of linearly independent Chen iterated integrals~\cite{ChenSymbol}. As we will argue in the main text, we expect that imposing these two conditions leads to a natural extension of the definition of \emph{canonical bases} beyond the polylogarithmic case. Thus, we will refer to our $\epsilon$-factorized basis as canonical to stress it admits these additional properties. 

The remainder of this paper is organized as follows. In section~\ref{sec:def}, we provide the setup of our calculation, including a description of the integral families in~\ref{sec:fam} and of the properties of the periods of the underlying K3 geometry in~\ref{sec:geo}. We continue in section~\ref{sec:diff}, where we describe the canonical system of differential equations. Its local solution in terms of iterated integrals close to $p^2=0$ is discussed in section~\ref{sec:sol}, while the use of these results to confirm the well-known three-loop wave function and charge renormalization constants is provided in section~\ref{sec:ren}. We finally describe the analytic continuation of our solution to cover the whole kinematic space in section~\ref{sec:analytic cont} and provide our conclusions in section~\ref{sec:conc}.

\section{Definitions and setup}
\label{sec:def}
We consider QED with a single fermion of (bare) mass $m$ and (bare) charge $-e$, which we will refer to as an electron. For later use, we introduce the electric coupling $\alpha = e^2/(4 \pi)$. Furthermore, we work in $D$ space-time dimensions and define the dimensional regulator $\epsilon$ by $D=D_0-2\epsilon$, where for our purposes, $D_0$ will be either two or four.

We briefly summarize some standard material on the free photon propagator, the full propagator including interactions, and the self-energy. A more detailed but similar treatment can be found, for example, in \cite{Grozin:2005yg}.
The photon propagator and the electron propagator in a general $R_\xi$ gauge are given by
\begin{align}
D_{\gamma, \text{free}}^{\mu \nu}(p) = -\frac{i}{p^2} \left(g^{\mu\nu}-(1-\xi) \frac{p^\mu p^\nu}{p^2}\right)\,,
\qquad
D_{e, \text{free}}(p) = \frac{i}{\slashed{p}-m}\,.
\end{align}
We are interested in computing the QED corrections to the
photon propagator. We denote by $i\, \Pi_{\text{bare}}^{\mu\nu}$ 
the sum of the corresponding one-particle-irreducible (1PI) diagrams 
and call $\Pi_{\text{bare}}^{\mu\nu}$ the bare photon self-energy.
As a consequence of the Ward identity, the self-energy of the photon is transverse, such that we can write
\begin{align}
    \Pi_{\text{bare}}^{\mu\nu}(p,m) = \left(p^2g^{\mu\nu}-p^\mu p^\nu\right) \Pi\left(p,m\right),
\end{align}
with one single, scalar form factor $\Pi\left(p^2\right)$.
We then obtain the full, bare propagator $D_{\gamma,\, \text{bare}}^{\mu\nu}$ by resumming the geometric series of 1PI diagrams,
    \begin{align}\label{eq:photon prop}
        D_{\gamma,\, \text{bare}}^{\mu\nu} (p,m) = -\frac{i}{p^2} \left[\frac{1}{1-\Pi(p,m)}\left( g^{\mu\nu}-\frac{p^\mu p^\nu}{p^2}\right) + \xi \frac{p^\mu p^\nu}{p^2}\right] \,.
    \end{align}
Since the self-energy is transverse, only the transverse part of the propagator receives loop corrections. Furthermore, this implies that the form factor $\Pi\left(p^2\right)$ is independent of the gauge parameter $\xi$.

Computing loop corrections to the full photon propagator then amount to computing $\Pi\left(p,m\right)$, which we extract by acting 
on the corresponding Feynman diagrams with a transverse projector,
\begin{align}\label{eq:transverse projection}
    \Pi\left(p,m\right) = \frac{1}{p^2(D-1)}\left(g_{\mu\nu}-\frac{p^\mu p^\nu}{p^2}\right)\Pi_{\text{bare}}^{\mu\nu}(p,m)\,.
\end{align}
Following the conventions in~\cite{Melnikov:2000zc,Duhr:2024bzt}, we expand the form factor in the coupling as
\begin{align}\label{eq:form factor expansion}
    \Pi (p,m) \coloneqq \sum\limits_{\ell=0}^\infty \left(\frac{\alpha}{\pi} C(\epsilon)\right)^{\ell} \Pi^{(\ell)}(p,m) \, ,
\end{align}
where $\ell$ is the loop order and 
\begin{align}\label{eq:definition C}
    C(\epsilon)= \Gamma(1+\epsilon) (4\pi)^\epsilon m^{-2\epsilon}\,.
\end{align}
We introduce this $\epsilon$-dependent prefactor to absorb some artifacts of dimensional regularization. Note that $C(0)=1$. 
Finally, we expand the coefficients $\Pi^{(\ell)}$ in the dimensional regulator $\epsilon$,
\begin{align}\label{eq:pi eps expansion}
    \Pi^{(\ell)}(p,m) \coloneqq \sum\limits_{n=-m}^{\infty} \epsilon^n\, \Pi^{(\ell, n)}(p,m)\,.
\end{align}
Since the scalar quantities $\Pi^{(\ell, n)}(p,m)$ are dimensionless, they depend on $p$ and $m$ only through the dimensionless ratio
\begin{align}
    x = \frac{p^2}{m^2}\,. \label{eq:defx}
\end{align}

\subsection{Feynman diagrams and integral families}
\label{sec:fam}
\begin{figure}[h]
\begin{center}
	\includegraphics[width=12cm]{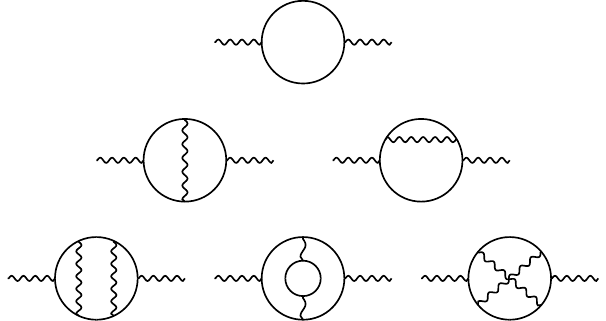}
\caption{Some examples of Feynman diagrams contributing to the QED photon self-energy up to three loops. In total, there are one, three, and 20 diagrams at one, two, and three loops, respectively.}
\label{fig:feyn diagrams}
\end{center}
\end{figure}

We generate the Feynman diagrams contributing to the photon self-energy up to three loops with \texttt{QGRAF}~\cite{Nogueira:1991ex}.
There is 1 diagram at one loop, 3 at two loops, and 20 at three loops. 
We show some examples in Fig.~\ref{fig:feyn diagrams}.
After applying the projector in eq.~\eqref{eq:transverse projection},
we perform the spinor and Dirac algebra to extract the scalar form factor $\Pi(p,m)$ using \texttt{FORM}~\cite{Vermaseren:2000nd,Kuipers:2012rf,Ruijl:2017dtg}. 
Since the one-loop and two-loop computations are purely polylogarithmic and straightforward, we focus our discussion on the three-loop contribution. Still, we also provide the results at lower-loop orders for completeness.

In order to map all Feynman integrals contributing to the photon self-energy at three loops to integral families, we consider three different families defined by
\begin{equation}
I^\text{fam}_{n_1 \, n_2 \, n_3 \, n_4 \, n_5 \, n_6 \, n_7 \, n_8 \, n_9}
= e^{3 \epsilon \gamma} \int \prod_{\ell=1}^3 \frac{\mathrm d^D k_\ell}{i \pi^{D/2}} \frac{1}{D_1^{n_1} \hdots D_9^{n_9}} \, ,
\end{equation}
where $\gamma$ is the Euler Mascheroni constant and the label `$\text{fam}$' can
take the three values $\text{fam} = A,B,C$.
The propagators for the three families are defined in Table~\ref{tab:props}. We divide each family into sectors, where a sector consists of all integrals with the same \textit{sector ID},
\begin{align}\label{eq:sec id}
    \text{ID} = \sum\limits_{j \,\text{with}\,n_j>0} 2^{j-1}\,.
\end{align}
\begin{table}[H]
\begin{center}
\begin{tabular}{| m{2.5cm} || m{3.5cm} | m{3.5cm} | m{3.5cm} |} 
 \hline
 Denominator & Integral family A & Integral family B & Integral family C  \\[5pt]
 \hline
$D_1$ & $k_1^2-m^2$         & $k_1^2-m^2$           & $k_1^2-m^2$           \\
$D_2$ & $k_2^2-m^2$         & $k_2^2-m^2$           & $k_2^2-m^2$           \\
$D_3$ & $k_3^2 -m^2$        & $k_3^2 $              & $k_3^2 $              \\
$D_4$ & $(k_1-p)^2 -m^2$    & $(k_1-p)^2 -m^2$      & $(k_1-p)^2 -m^2$      \\
$D_5$ & $(k_2-p)^2 -m^2 $   & $(k_2-p)^2 -m^2 $     & $(k_2-p)^2 -m^2 $     \\
$D_6$ & $(k_3-p)^2 -m^2 $   & $(k_3-p)^2 $          & $(k_3-p)^2 $          \\
$D_7$ & $(k_1-k_2)^2$       & $(k_1-k_3)^2 -m^2$    & $(k_1-k_2)^2$         \\
$D_8$ & $(k_1-k_3)^2$       & $(k_2-k_3)^2 -m^2$    & $(k_2+k_3-p)^2-m^2$   \\
$D_9$ & $(k_2-k_3)^2$       & $(k_1+k_2-k_3-p)^2$   & $(k_1+k_3-p)^2-m^2$   \\
 \hline
\end{tabular}
\caption{Definitions of the propagators of the three scalar integral families $A, B, C$.}
\label{tab:props}
\end{center}
\end{table}
In the following, we label sectors by their ID. Out of all sectors we need to compute in a given family, we refer to the one with the highest number of different propagators in the denominator as the \textit{top sector} of that family.
All scalar integrals can be reduced to masters using symmetries and integration-by-parts (IBP) relations~\cite{Tkachov:1981wb, Chetyrkin:1981qh, Laporta:2004rb}.
In this way, it is easy to see that family $C$ does not introduce any additional master integrals with respect to those generated by family $A$ and $B$. 
To perform the reduction we have employed \texttt{Reduze 2}~\cite{Studerus:2009ye,vonManteuffel:2012np} and \texttt{Kira 2}~\cite{Maierhofer:2017gsa,Klappert:2019emp,Klappert:2020nbg}. We find 36 independent master integrals in total, where the first 20 belong to family $A$, and the last 16 come from family $B$. 
As a convenient choice of initial Laporta integrals, we follow the general prescriptions provided in~\cite{Gorges:2023zgv} and consider the basis
\allowdisplaybreaks
\begin{align}
&I^A_{1 1 1 0 0 0 0 0 0} \,, \qquad 
&I^A_{1 1 1 1 0 0 0 0 0} \,, \qquad 
&I^A_{0 1 1 1 0 0 1 0 0} \,, \qquad 
&I^A_{-1 1 1 1 0 0 1 0 0} \,, \qquad 
&I^A_{0 1 1 0 0 0 1 1 0} \,, \nonumber \\ 
&I^A_{0 0 1 0 1 0 1 1 0} \,, \qquad 
&I^A_{-1 0 1 0 1 0 1 1 0} \,, \qquad 
&I^A_{0 -1 1 0 1 0 1 1 0} \,, \qquad
&I^A_{1 1 1 1 1 0 0 0 0} \,, \qquad 
&I^A_{0 1 1 1 0 1 1 0 0} \,, \nonumber \\ 
&I^A_{-1 1 1 1 0 1 1 0 0} \,, \qquad
&I^A_{-1 1 1 1 0 0 1 1 0} \,, \qquad 
&I^A_{-2 1 1 1 0 0 1 1 0} \,, \qquad 
&I^A_{0 1 1 1 0 0 1 1 -1} \,, \qquad 
&I^A_{0 1 1 0 1 0 1 1 -1} \,, \nonumber \\
&I^A_{1 1 1 1 1 1 0 0 0} \,, \qquad 
&I^A_{0 1 2 1 1 0 1 1 0} \,, \qquad
&I^A_{0 1 1 0 1 1 2 1 0} \,, \qquad 
&I^A_{1 1 1 0 1 1 1 1 0} \,, \qquad 
&I^A_{1 1 1 1 1 1 1 1 0} \,, \nonumber  \\
&I^B_{1 1 0 0 0 0 1 1 0} \,, \qquad 
&I^B_{0 1 0 1 0 0 1 1 0} \,, \qquad
&I^B_{0 2 0 1 0 0 1 1 0} \,, \qquad 
&I^B_{0 3 0 1 0 0 1 1 0} \,, \qquad 
&I^B_{1 1 0 1 0 0 1 1 0} \,, \nonumber  \\ 
&I^B_{0 0 1 1 1 0 1 1 0} \,, \qquad
&I^B_{-1 0 1 1 1 0 1 1 0} \,,\qquad 
&I^B_{1 1 0 1 1 0 2 1 0} \,, \qquad
&I^B_{2 1 0 1 1 0 2 1 0} \,, \qquad 
&I^B_{0 1 1 2 0 1 1 0} \,,   \nonumber  \\ 
&I^B_{0 1 1 2 0 1 1 0} \,,   \qquad 
&I^B_{0 1 1 1 0 0 2 1 1} \,, \qquad
&I^B_{0 1 1 1 0 0 1 2 1} \,, \qquad 
&I^B_{0 1 1 1 0 0 3 1 1} \,, \qquad 
&I^B_{1 1 1 1 1 0 1 1 1} \,, \nonumber  \\  
&I^B_{1 1 1 1 1 -1 1 1 1} \,.\qquad
& \, \qquad 
& \, \qquad
& \, \qquad 
& \, \nonumber   
\label{eq:misI}
\end{align}

\begin{figure}[h]
\begin{center}
	\includegraphics[width=10cm]{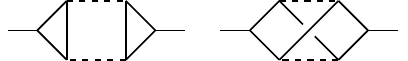}
\caption{Planar and non-planar top sectors of family A and B that admit a fully massive four-particle cut.
}
\label{fig:top sectors}
\end{center}
\end{figure}

As we discussed earlier, the photon self-energy at one and two loops is purely polylogarithmic, unlike the electron self-energy where elliptic integrals already appear at two loops~\cite{Duhr:2024bzt}. 
In Fig.~\ref{fig:top sectors}, we draw the planar and non-planar top sectors contributing to the three-loop photon self-energy.
As it is easy to see, they both involve diagrams with four massive particle cuts, as opposed to the electron self-energy at three loops~\cite{Duhr:2024bzt}. This cut is related to the three-loop equal-mass banana integral, which is known to be associated with a K3 surface~\cite{Primo:2017ipr,Broedel:2019kmn,Bonisch:2020qmm}. 
Indeed, by an analysis of the relevant maximal cuts~\cite{Primo:2016ebd,Primo:2017ipr} in Baikov representation~\cite{Baikov:1996iu,Baikov:1996rk,Frellesvig:2017aai}, one can
easily prove that all higher sectors beyond the banana do not introduce any further geometry beyond the K3 associated to the banana integral.

Before describing our canonical basis for this set of master integrals, it is useful to recall some well-known properties of the associated K3 geometry.

\subsection{The K3 geometry}
\label{sec:geo}

The maximal cut of the three-loop equal-mass banana graph in $D=2$ dimensions satisfies a third-order Picard-Fuchs differential equation~\cite{Primo:2017ipr}
\begin{align}
    \mathcal{L}_3(x) \varpi_i(x) &=0 \qquad \text{for } i=0,1,2,
\end{align}
with
\begin{align}
   \mathcal{L}_3(x) &= \left(x\dv{x}\right)^3 +\frac{3 x (x-10)}{(x-16) (x-4)}\left(x\dv{x}\right)^2
    +\frac{3 x(x-6)}{(x-16) (x-4)}\left(x\dv{x}\right)+\frac{x}{(x-16)}\,.
\label{eq:pfK3}
\end{align}
This differential equation has regular singular points at $x\in \{0,4,16,\infty \}$, and its solutions are the three periods of a one-parameter K3 surface.
In the following, we give our choice of local solutions around these four singular points. We label our solutions by $\varpi_i^{[p]}$ for $i=0,1,2$, where $p=\{0,4,16,\infty\}$ refers to the expansion point. Moreover, we also consider the solution close to the point $x=1$. Even though it is not a singular point of eq. \eqref{eq:pfK3}, it will become important in the context of analytic continuation in \cref{sec:analytic cont}, since it is a singular point of the differential equation satisfied by the master integrals.
\begin{itemize}
    \item Expanding at $x=0^+$, we choose 
    our three independent solutions as
    \begin{align}\label{eq:period expansions 0}
        \varpi_0^{[0]}(x) &= 1+\frac{x}{16}+\frac{7 x^2}{1024}+\frac{x^3}{1024}+\frac{679 x^4}{4194304}+ \mathcal{O}(x^5) \, ,\nonumber\\
        \varpi_1^{[0]}(x) &= \varpi_0^{[0]}(x)\log(x)+\Sigma_1^{[0]}(x) \, ,\\
        \varpi_2^{[0]}(x) &= \frac{1}{2}\varpi_0^{[0]}(x)\log^2(x)+\Sigma_1^{[0]}(x)\log(x)+\Sigma_2^{[0]}(x) \, , \nonumber
    \end{align}
    where
    \begin{equation}
        \begin{split}
            \Sigma_1^{[0]}(x) &=\frac{3 x}{32}+\frac{57 x^2}{4096}+\frac{73 x^3}{32768}+\frac{13081 x^4}{33554432} + \mathcal{O}(x^5) \, , \\
            \Sigma_2^{[0]}(x) &= \frac{9 x^2}{2048}+\frac{135 x^3}{131072}+\frac{7089 x^4}{33554432}+ \mathcal{O}(x^5) \, .
        \end{split}
    \end{equation}
    We notice that $x=0$ is a point of Maximal Unipotent Monodromy (MUM), as it can be seen from the fact that the complete tower of logarithms up to $\log^2{(x)}$ is required to describe the homogeneous solutions locally close to this point.

    \item For $x=1$, we use the local variable $u=x-1$. Since this is not a singular point, all solutions are regular, and we order them by the starting power of their Taylor expansion at $u=0$,
    \begin{align}
        \varpi_0^{[1]}(u) &= 1+\frac{7 u^2}{180}-\frac{u^3}{50}+\frac{541 u^4}{32400} + \mathcal{O}(u^5) \, ,\nonumber\\
        \varpi_1^{[1]}(u) &= u-\frac{u^3}{90}+\frac{23 u^4}{360} + \mathcal{O}(u^5) \, ,\\
        \varpi_2^{[1]}(u) &= u^2-\frac{4 u^3}{5}+\frac{133 u^4}{180} + \mathcal{O}(u^5) \, . \nonumber
    \end{align}

    \item For $x=4^+$, we use the local variable $v=x-4$ and find that the homogeneous solutions have an algebraic branching point but no logarithms. Therefore, this is not a MUM point. We choose our solutions as
    \begin{align}\label{eq:varpiv}
        \varpi_0^{[4]}(v) &= 1-\frac{v}{12}+\frac{v^2}{72}-\frac{11 v^3}{4320}+\frac{53 v^4}{103680} + \mathcal{O}(v^5) \, ,\nonumber\\
        \varpi_1^{[4]}(v) &= v-\frac{v^2}{6}+\frac{5 v^3}{144}-\frac{11 v^4}{1512} + \mathcal{O}(v^5) \, ,\\
        \varpi_2^{[4]}(v) &= \sqrt{v} \left(1-\frac{v}{8}+\frac{3 v^2}{128}-\frac{493 v^3}{107520}+\frac{1097 v^4}{1146880}+ \mathcal{O}(v^5)\right)  \, . \nonumber
    \end{align}

    \item At $x=16^+$, the situation is very similar as close to $x=4^+$. We use the local variable $w=x-16$ and define our local solutions as
    \begin{align}
        \varpi_0^{[16]}(w) &= 1-\frac{w}{24}+\frac{5 w^2}{2304}-\frac{17 w^3}{138240}+\frac{97 w^4}{13271040} + \mathcal{O}(w^5) \, ,\nonumber\\
        \varpi_1^{[16]}(w) &= w-\frac{w^2}{12}+\frac{7 w^3}{1152}-\frac{83 w^4}{193536} + \mathcal{O}(w^5) \, ,\\
        \varpi_2^{[16]}(w) &= \sqrt{w} \left(1-\frac{w}{16}+\frac{w^2}{256}-\frac{107 w^3}{430080}+\frac{37 w^4}{2293760} + \mathcal{O}(w^5)\right) \, . \nonumber
    \end{align}
    
    \item Finally, $x=+\infty$ turns out to be another MUM point. Here
    we use $z=1/x$ as the local variable and choose the three solutions ordered by powers of logarithms as
    \begin{align}
        \varpi_0^{[\infty]}(z) &=z+4 z^2+28 z^3+256 z^4+ \mathcal{O}(z^5) \, ,\nonumber\\
        \varpi_1^{[\infty]}(z) &= \varpi_0^{[\infty]}(z)\log(z)+\Sigma_1^{[\infty]}(z) \, ,\\
        \varpi_2^{[\infty]}(z) &= \frac{1}{2}\varpi_0^{[\infty]}(z)\log^2(z)+\Sigma_1^{[\infty]}(z)\log(z)+\Sigma_2^{[\infty]}(z) \, , \nonumber
    \end{align}
    where
    \begin{equation}
        \begin{split}
            \Sigma_1^{[\infty]}(z) &= 6 z^2+57 z^3+584 z^4+ \mathcal{O}(z^5) \, , \\
            \Sigma_2^{[\infty]}(z) &= 18 z^3+270 z^4 + \mathcal{O}(z^5) \, .
        \end{split}
    \end{equation}
\end{itemize}
We stress here that at the MUM points $x=0$ and $x=\infty$, we have chosen $\varpi_0^{[p]}$ to be the unique (up to normalization) holomorphic solution. This object can be seen as the leading singularity associated with the corner integral\thinspace\footnote{By the \textit{corner integral} of a sector, we refer to the (unique) member of that sector where all propagator powers $n_i$ are either zero or one.} in the three-loop banana family in $D=2$. In this sense, it generalizes
an algebraic function in the polylogarithmic case or the elliptic integral of the first kind for a one-dimensional elliptic curve.

As we saw above, around non-MUM regular singular points, one finds, in
general, more regular solutions, such that there is no unique choice of a
holomorphic period $\varpi_0^{[p]}$. Since the choice of $\varpi_0^{[p]}$ is central for the construction of a canonical basis, we will elaborate more in detail on the specific choices we made for $\varpi_0^{[p]}$ around non-MUM points in \cref{sec:analytic cont}.

The full set of periods, together with their first and second derivatives, can be arranged in the so-called Wronskian matrix
\begin{align}
     W &= \begin{pNiceMatrix}
        \varpi_0(x) & \varpi_1(x) & \varpi_2(x) \\
        \varpi_0'(x) & \varpi_1'(x) & \varpi_2'(x) \\
        \varpi_0''(x) & \varpi_1''(x) & \varpi_2''(x)
        \end{pNiceMatrix}\,.
\end{align}
By construction, this is the matrix of solutions to the third-order system of differential equations satisfied by the corner integral of the banana sector and its two $ x$ derivatives.
Importantly, the periods and their derivatives are not independent but, instead fulfill a set of quadratic relations, which generalizes the famous Legendre relation for elliptic integrals.
These quadratic relations in the case of general Calabi-Yau geometries are referred to by Griffiths transversality conditions~\cite{Bonisch:2021yfw,Gorges:2023zgv}.
In the case of the K3 geometry associated with the equal-mass banana, they can be expressed compactly in matrix form as
\begin{align}
    Z = W \,\Sigma \, W^T \,, \qquad \mbox{with}
    \qquad  \Sigma = \begin{pNiceMatrix}
        0 & 0 & 1 \\
        0 & -1 & 0 \\
        1 & 0 & 0
        \end{pNiceMatrix},
\end{align}
and
\begin{equation}
     \begin{split}
        Z^{-1} &= \begin{pNiceMatrix}
            (x-8) x & 2 x \left(x^2-15 x+32\right) & x^2 \left(x^2-20 x+64\right) \\
            2 x \left(x^2-15 x+32\right) & -x^2 \left(x^2-20 x+64\right) & 0 \\
            x^2 \left(x^2-20 x+64\right) & 0 & 0 
        \end{pNiceMatrix}.
    \end{split}
\end{equation}

We can solve these relations for the second derivative
of the holomorphic period, yielding
\begin{align}\label{eq:griffiths}
   \varpi_0''(x) = \frac{1}{2}\left(-\frac{(x-8)}{ (x-16) (x-4) x} \varpi_0(x) -\frac{4 \left(x^2-15 x+32\right) }{(x-16) (x-4) x} \varpi_0'(x) +\frac{\varpi_0'(x)^2}{ \varpi_0(x)}\right),
\end{align}
which we will use to obtain and simplify the canonical basis of differential equations.

\subsubsection*{The K3 differential operator as a symmetric square}

Even though it is not essential to our procedure, we recall here that the differential operator of the K3 geometry defined in eq.~\eqref{eq:pfK3} is known to be the symmetric square~\cite{Joyce1,Primo:2017ipr,Broedel:2019hyg} of the second order differential operator
\begin{align}
    \mathcal{L}_2(x) \coloneqq \left(x\dv{x}\right)^2 +\frac{ x (x-10)}{(x-16) (x-4)}\left(x\dv{x}\right) +\frac{x(x-8)}{4 (x-16) (x-4)}\,.
\end{align}
It is a very general result that every Picard-Fuchs differential operator of a one-parameter K3 surface is a symmetric square of a second-order operator~\cite{doran,BognerThesis,Bogner:2013kvr}.
From a practical point of view, this implies that, 
if we define $\pi_0(x), \pi_1(x)$ to be the solutions of
\begin{align}
    \mathcal{L}_2(x) \pi_i(x) =0 \qquad \text{for } i=0,1,
\end{align}
then the three solutions $\varpi_i(x)$ of eq.~\eqref{eq:pfK3} can be obtained by taking the symmetric products
\begin{align}\label{eq:sym square relations}
    \varpi_0(x) = \pi_0^2(x)\,, \qquad \varpi_1(x) = \pi_0(x)\pi_1(x)\,, \qquad \varpi_2(x) = \frac{1}{2} \pi_1^2(x)
\end{align}
of these two periods, where we choose a convenient normalization. 
These relations can easily be checked explicitly using, for example, the definition of the periods in terms of their series expansions close to $x=0^+$,
\begin{equation}
    \begin{split}
        \pi_0(x)&=1+\frac{x}{32}+\frac{3 x^2}{1024}+\frac{13 x^3}{32768}+\frac{539 x^4}{8388608}+ \mathcal{O}(x^5)\, , \\
        \pi_1(x)&=\pi_0(x) \log(x)+\frac{3 x}{32}+\frac{45 x^2}{4096}+\frac{211 x^3}{131072}+\frac{9065 x^4}{33554432}+ \mathcal{O}(x^5)\,.
    \end{split}
\end{equation}
Using these expressions, one can check that they indeed reproduce the expansions in eq.~\eqref{eq:period expansions 0} according to the relations in eq.~\eqref{eq:sym square relations}. Note that this symmetry of the homogeneous solutions of the three-loop banana graph is very special, and it is not expected to hold for arbitrary values of the propagator masses. 

\section{Differential equations and $\epsilon$-factorization}
\label{sec:diff}

Now that we have described the properties of the geometry involved, we have all the ingredients to consider the analytic calculation of the master integrals relevant to the three-loop photon self-energy with the differential equations method.
Using IBP identities, we derive a system of differential equations
\begin{align}
    \dv{\vec{I}}{x} = A(x;\epsilon) \vec{I} \, ,
\end{align}
satisfied by the set of master integrals, which we denote by $\vec{I}$. It is convenient to order the master integrals based on the increasing number of propagators, i.e., each master integral appears in $\vec{I}$ below all its subtopologies. In this way, the matrix $A(x,\epsilon)$ takes a lower block-triangular form. Schematically we find
\begin{align}
     A(x;\epsilon)=\scalemath{0.7}{
     \begin{pNiceMatrix}[columns-width = 0.4cm]
        \Block[borders={bottom,right}]{1-1}{*}  &  & & & & & & & & \\
         &  \Block[draw]{1-1}{*}& & & & & \Block{3-3}<\LARGE>{0}& & &\\
         &  & \Block[draw]{2-2}<\Large>{*}& & & & & & & \\
         &  & & & & & & & & \\
         &  & & & \Block[draw]{1-1}{*} & & & & & \\
         &  & & & &\Block[draw]{3-3}<\LARGE>{*} & & & & \\
         &\Block{3-3}<\LARGE>{*}  & & & & & & & & \\
         &  & & & & & & & & \\
         &  & & & & & & &\Block[borders={top,left}]{1-1}{\ddots}  & \\
         &  & & & & & & & &\Block[borders={top,left}]{1-1}{*} 
        \end{pNiceMatrix}
        }\,.
\end{align}
As the next step, we may perform a rotation in the space of
master integrals defining the new basis
\begin{align}
    \vec{J} = T(x;\epsilon) \vec{I}\,,
\end{align}
such that the new differential equation reads
\begin{align}
    \dv{\vec{J}}{x}= B(x;\epsilon) \vec{J}\,, \qquad \text{where}\qquad B(x;\epsilon)= T A T^{-1} + \dv{T}{x} T^{-1}\,.
\end{align}
To solve the differential equation as a series expansion in $\epsilon$, our strategy is to find a matrix $T(x;\epsilon)$ such that
the system becomes $\epsilon$-factorized,
\begin{align}
    B(x;\epsilon) = \epsilon\, G(x)\, .
\label{eq:epsfact}
\end{align}

We stress here an important point. As mentioned in~\cref{sec:intro}, there is a fundamental difference between a randomly chosen $\epsilon$-factorized system and a proper generalization of a \emph{canonical basis}~\cite{Henn:2013pwa}. 
In fact, as it is well known, also in the purely polylogarithmic case, it is possible to find $\epsilon$-factorized bases whose elements are neither of uniform transcendental weight (UT) nor pure.
In the polylogarithmic case, UT and purity are guaranteed if one obtains a set of differential equations with logarithmic singularities (single poles) and whose boundary conditions are UT. This, in turn, can conjecturally be realized by choosing integrals with unit leading singularities~\cite{ArkaniHamed:2010gh, Henn:2013pwa, Henn:2020lye}.
As it was explained in~\cite{Gorges:2023zgv}, this construction can be generalized to non-trivial geometries starting from the fact that, compared to the purely polylogarithmic case, differential forms with simple poles on an arbitrary geometry (which are in general referred to as differential forms of the third kind) are not sufficient to span its full cohomology. Instead, new integrals are required with either no poles (forms of the first kind) or double and higher poles (forms of the second kind). The forms of the first kind
are mapped to the periods of the underlying manifold.
Interestingly, canonical Feynman integrals corresponding to the forms of the first kind can typically be chosen through a similar residue analysis, while a convenient starting basis for the  integrals of the second kind 
can be obtained by taking enough derivatives of the corresponding first-kind integrals.
The required integrand analysis is usually performed in a parametric representation for the Feynman integrals, as the Feynman-Schwinger or the Baikov representation~\cite{Baikov:1996iu, Baikov:1996rk}.
In addition, the existence of forms of the second kind requires a further step in constructing a canonical basis. This can be best understood considering the period matrix locally close to a MUM point where the full tower of logarithms is present.
Here, one can split the Wronskian into a semi-simple and a unipotent part, rotate away the semisimple part, and rescale the resulting new basis by appropriate powers of $\epsilon$.
These steps effectively separate the components of the master integrals that have different transcendental weights and adjust their normalization accordingly.
One can then consistently extend this construction to any other kinematic point (regular or singular), as explained in section~\ref{sec:analytic cont}.
More details about this procedure for elliptic and K3 geometries are provided in~\cite{Gorges:2023zgv}, while its generalization to Calabi-Yau geometries will be described elsewhere~\cite{CYeps}.

Importantly, once the procedure above has been employed, one is left with
a system of differential equations whose differential forms have conjecturally at most single poles close to each regular singular point and whose independence can easily be checked, for example, using their series expansion representation.\footnote{Proving the independence of the differential forms in full generality is a delicate problem since one cannot easily exclude non-trivial relations among the differential forms which amount to a total derivative (or an exact form) of some other function. Here, we only exclude relations that involve a total differential of an arbitrary linear combination of the forms themselves.} We expect that imposing these two additional conditions on $\epsilon$-factorized bases leads to a natural extension of the definition of \emph{canonical bases} beyond the polylogarithmic case.\footnote{It will be interesting in the future to elucidate the possible equivalence between these requirements, and the concept of a C-form introduced in~\cite{Duhr:2024xsy}. We thank C. Duhr for highlighting this connection to us.} The explicit rotation for the banana sector and our $\epsilon$-factorized canonical basis choice are provided in \cref{app:canbasis}.

Going back to eq.~\eqref{eq:epsfact}, it is then convenient to write $G(x)$ in terms of a set of the independent differential forms $\omega_i = f_i(x)\mathrm dx$ as 
\begin{align}
    G(x) = \sum\limits_i G_i\, f_i(x) \, ,
\end{align}
and cast the differential equation into the form
\begin{align}
    \dd \vec{J} =  \epsilon \,\left(\sum\limits_i G_i\, \omega_i\right) \vec{J} \, .
\end{align}
From here, we may integrate the differential equation iteratively, order by order in $\epsilon$. Since the photon self-energy depends only on one dimensionless variable, the solutions to the differential equation can be written in terms of (Chen) iterated integrals~\cite{ChenSymbol} of the form
\begin{align}
    I(f_{i_n},\hdots,f_{i_1};x) = \int_{x_0}^x \mathrm dx_n\, f_{i_n}(x_n) \hdots \int_{x_0}^{x_3} \mathrm dx_2\, f_{i_2}(x_2) \int_{x_0}^{x_2} \mathrm dx_1 \,f_{i_1}(x_1)
\end{align}
with boundary point at $x=x_0$. In our specific case, we work in the region $0<x<4$ for definiteness and with $x_0=0$ as our boundary point.

In deriving a canonical basis for our problem, we are forced to introduce three square roots
\begin{align}\label{eq:def square roots}
    r_1(x)=\sqrt{x}\,,\quad r_2(x)=\sqrt{4-x}\,,\quad r_3(x)=\sqrt{16-x} \, ,
\end{align}
which we choose to be real in the region $0<x<4$ under consideration.\footnote{In fact, the three square roots appear only in the two combinations $r_1\, r_2$ and $r_2\, r_3$, yet we choose to define them as three separate square roots of linear polynomials. This will be convenient for the analytic continuation discussed in \cref{sec:analytic cont}.} With these, the final differential equation may be written in terms of 24 differential forms
    \begin{align}\label{eq:dlog kernels}
    f_i \in \Bigg\{&\frac{1}{x},
    \quad \frac{1}{x-1},
    \quad \frac{1}{x-4},
    \quad \frac{1}{r_1(x) r_2(x)},
    \quad \frac{1}{16-x}\Bigg\} \quad \text{for } i=1,\dots,5,\quad \text{and}  \\
    f_i \in \Bigg\{&\frac{1}{r_2(x) r_3(x) x \varpi_0(x)},
    \frac{G_2(x) }{r_2(x) r_3(x) x \varpi_0(x)}, 
    \frac{G_2(x)^2}{r_2(x) r_3(x) x \varpi_0(x)},
    \frac{G_2(x)^3}{r_2(x) r_3(x) x \varpi_0(x)}, \nonumber\\
    &\frac{G_3(x)}{r_2(x) r_3(x) x \varpi_0(x)},
    \frac{G_2(x) G_3(x) }{r_2(x) r_3(x) x \varpi_0(x)},
    \frac{G_3(x) }{r_1(x) r_2(x)},
    \frac{G_3(x)}{x},
    \frac{G_3(x)}{4-x},
    \frac{G_3(x)}{16-x},\nonumber\\
    &\varpi_0(x),
    \frac{\varpi_0(x)}{x},
    \frac{\varpi_0(x)}{4-x},
    \frac{r_1(x) r_2(x) \varpi_0(x)}{x},
    \frac{r_1(x) \varpi_0(x)}{r_2(x)},
    \frac{r_1(x) \varpi_0(x)}{r_2(x)(4-x)},\nonumber\\
    &\frac{(x+8)^2 \left(x^2-8 x+64\right)   \varpi_0(x)}{r_2(x) r_3(x) (16-x) (4-x) x},
    \frac{(x+8)^2 \left(x^2-8 x+64\right) G_2(x) \varpi_0(x)}{r_2(x) r_3(x) (16-x) (4-x) x},\nonumber\\
    &\frac{(x-8) (x+8)^3 \varpi_0(x)^2}{(16-x)^2 (4-x)^2}\Bigg\} \quad \text{for } i=6,\dots,24.\label{eq:elliptic kernels}
    \end{align}
The first five of these differential forms are dlog-forms, and in particular, $\omega_4$ is the classical logarithmic differential form which already appeared in the calculation of the one-loop massive bubble,
\begin{align}
     \omega_4= i \,\text{dlog} \left(\frac{\sqrt{4-s}-i \sqrt{s}}{\sqrt{4-s}+i \sqrt{s}}\right) \, .
\end{align}
The remaining 19 differential forms involve kernels expressed through powers of the holomorphic K3 period $\varpi_0(x)$ and the additional functions
\begin{align}\label{eq:def G functions}
G_1(x) &\coloneqq -\int\limits_0^x \mathrm d u \frac{(u-8) (u+8)^3 \varpi_0(u)^2}{32 (u-16)^2 (u-4)^2} = \frac{x}{32}+\frac{x^2}{64}+\frac{93 x^3}{16384}+\frac{1841 x^4}{1048576} + \mathcal O(x^5)\,,\nonumber\\
    G_2(x) &\coloneqq \int\limits_0^x \mathrm d u \frac{8 G_1(u) r_3(u) r_2(u)}{(4-u) (16-u) u \varpi_0(u)} = \frac{x}{32}+\frac{19 x^2}{2048}+\frac{167 x^3}{65536}+\frac{5539 x^4}{8388608} + \mathcal O(x^5)\, , \nonumber\\
    G_3(x) &\coloneqq \int\limits_0^x \mathrm d u \frac{(u+2) r_1(u) r_2(u) \varpi_0(u)}{(u-4)^2 u}  \\
    &= \frac{\sqrt{x}}{2}+\frac{5 x^{3/2}}{32}+\frac{75 x^{5/2}}{2048}+\frac{139 x^{7/2}}{16384}+\frac{16591 x^{9/2}}{8388608} + \mathcal O(x^{11/2}) \, . \nonumber
\end{align}
Note that these functions involve only the holomorphic period $\varpi_0$. Moreover,  $G_1$ does not appear explicitly in the kernels listed above but implicitly in the definition of $G_2$.
We stress here that, at variance with the $D=2$ part of the three-loop banana graph, the extra singularities appearing in our kernels make it such that the integrals stemming from them cannot be straightforwardly expressed as iterated integrals over modular forms. This means that the space of functions required to
solve this problem is more general than the one required for
the three-loop banana graph~\cite{Broedel:2019kmn}.
As we will see in the next section, many of these kernels do not contribute to the $\mathcal{O}(\epsilon^0)$-part of the self-energy, i.e., the finite remainder.

At one and two loops, we observe that the canonical differential equations contain only a subset of the dlog kernels in~\eqref{eq:dlog kernels}. In particular, at one loop, only two kernels contribute
\begin{align}\label{eq:kernels 1 loop}
    f_i^{(1l)}\in \left\{\frac{1}{x-4},
    \quad \frac{1}{r_1(x) r_2(x)}\right\} \, ,
\end{align}
while at two loops, the list of kernels reads
\begin{align}\label{eq:kernels 2 loop}
    f_i^{(2l)}\in \left\{\frac{1}{x},
    \quad \frac{1}{x-4},
    \quad \frac{1}{r_1(x) r_2(x)}\right\} \, .
\end{align}

\section{Analytic solution to three loops}
\label{sec:sol}
The canonical differential equation immediately gives the general solution for our master integrals in terms of iterated integrals once the boundary values are known. We determine the boundary values as follows. Focussing again on the region $0<x<4$, we choose to fix the boundary constants at $x=0$. To do so, we follow the same procedure described, for example, in~\cite{Duhr:2024bzt} and compute them numerically to high precision using \texttt{AMFlow}~\cite{Liu:2017jxz, Liu:2022chg}. 
Then, using an integer relation algorithm \texttt{PSLQ}~\cite{Ferguson1998APT}, we match these numerical results to linear combinations of analytical transcendental numbers that we expect to appear. Of course, in general, it is not clear beforehand what set of constants may appear when evaluating the integrals at an arbitrary point. 
At $x=0$, however, the K3 geometry degenerates to a sphere, such that we expect the appearing transcendental numbers to be of polylogarithmic origin. The specific letters of the polylogarithms can be restricted, noticing that at $p^2=0$, the Feynman integrals reduce to the corresponding set of tadpole integrals, which contribute to the charge renormalization constant in QED.
At three loops, the required set of constants at $x=0$ turns out to be
\begin{align}
    \left\{\pi,\log(2),\zeta (3),\zeta (5),\text{Li}_4\!\left(\tfrac{1}{2}\right),\text{Li}_5\!\left(\tfrac{1}{2}\right) \right\} \, ,
\end{align}
and combinations thereof. This completes the computation of the self-energy in terms of formal iterated integrals in the region $0<x<4$. Analytic results can be found in the ancillary files published together with the arXiv submission of this paper. In the following section, we describe the general structure of the solution.

\subsection{Solution in terms of iterated integrals}

Similarly to what was recently observed in the calculation of the electron self-energy~\cite{Duhr:2024bzt}, the set of kernels appearing in the final result for the photon self-energy is a subset of the kernels that are present in the canonical differential equation (cf. eqs.~\eqref{eq:dlog kernels} and~\eqref{eq:elliptic kernels}). 
The absence of some of them is caused either by the insertion of the boundary conditions, which may set some contributions to zero, or by cancellations in the physical result, i.e., after insertion of the integrals into the full self-energy. 

Focussing on the three-loop contribution through order $\epsilon^0$, we find that it contains iterated integrals of kernels drawn from the much smaller set
\begin{align}
    \Bigg\{&\frac{1}{x},
    \frac{1}{x-1},
    \frac{1}{x-4},
    \frac{1}{r_1(x) r_2(x)},
    \frac{1}{r_2(x) r_3(x) x \varpi_0(x)},
    \frac{ G_3(x)}{r_2(x) r_3(x) x \varpi_0(x)},
    \frac{G_3(x)}{r_1(x) r_2(x)},\nonumber\\
    &
    \frac{G_3(x)}{x},
    \frac{G_3(x)}{4-x},
    \varpi_0(x),
    \frac{\varpi_0(x)}{x},
    \frac{\varpi_0(x)}{4-x},
    \frac{r_2(x) \varpi_0(x)}{r_1(x)},
    \frac{r_1(x) \varpi_0(x)}{r_2(x)},
    \frac{r_1(x) \varpi_0(x)}{r_2(x)(4-x)}\Bigg\} \, ,
\end{align}
i.e., it involves only 15 of the 24 kernels appearing in the differential equation. 
Interestingly, many of the K3 kernels do not contribute. 
In particular, we notice that neither kernels containing $\varpi_0^2(x)$ nor any power of $G_2(x)$ are present in the finite remainder. The absence of $\varpi_0^2(x)$ has also been observed in the elliptic case for the electron self-energy at two and three loops~\cite{Duhr:2024bzt}, where it was argued that this contribution comes from the second master integral of the banana sector (the third one in our case), at a too high order in $\epsilon$ to contribute to the finite remainder. This argument also seems to apply to the case at hand. However, it does not explain the absence of $G_2(x)$, which instead disappears from the finite part of the self-energy only through
a delicate cancellation among different master integrals (we recall here for the reader's convenience that $G_1(x)$ only appears in the matrix of differential equations indirectly through $G_2(x)$, but never alone).
We notice, in particular, that both $G_2(x)$ and $G_1(x)$ are also present in the inverse rotation from our canonical basis back to the initial one, but they still do not appear in the final result.
On the other hand, $G_3(x)$, which does not involve $\varpi_0^2(x)$, does appear in the kernels in all stages of the computation and, in particular, also in the finite remainder.

These findings suggest that the observed cancellations could have been made manifest at the master integrals level by performing a rotation in the space
of master integrals, for example, following the construction described in~\cite{Chicherin:2021dyp, Gehrmann:2024tds}. One of the cancellations one may want to make explicit is the one of the branch cut starting at $x=16$.
We do not pursue this transformation here. 
On the other hand, we find it particularly evocative to stress that both $G_1(x)$ and $G_2(x)$ are iterated integrals of kernels involving $\varpi_0^2(x)$ (cf. eq.\eqref{eq:def G functions}). 
This means that the complete cancellation of these kernels guarantees that
none of the iterated integrals appearing up to $\mathcal{O}(\epsilon^0)$
ever requires integrating over $\varpi_0^2(x)$ (cf. eq.\eqref{eq:def G functions}).
From the few analytic calculations involving geometries beyond the Riemann sphere available in the literature, the absence of iterated integrals over kernels that depend on higher powers of the holomorphic period appears to be a very general feature. In the future, it will be interesting to investigate whether these patterns hold in more examples and, if they do, 
if they can be explained in terms of physical requirements satisfied by the master integrals or by the final result. In this case, they could constitute an important ingredient for the generalization of the Boostrap program beyond polylogarithms, see for example~\cite{Morales:2022csr}.

For completeness, we also report on the kernels that appear in the final results at one and two loops. For the former, only one of the two kernels in eq.~\eqref{eq:kernels 1 loop}
\begin{align}\label{eq:kernels 1 loop prop}
    \Bigg\{\frac{1}{r_1(x) r_2(x)}\Bigg\}
\end{align}
contributes. At two loops, all three kernels in eq.~\eqref{eq:kernels 2 loop} that appear in the differential equation,
\begin{align}\label{eq:kernels 2 loop prop}
    \Bigg\{\frac{1}{x},
    \quad \frac{1}{x-4},
    \quad \frac{1}{r_1(x) r_2(x)}\Bigg\} \, ,
\end{align}
are also present in the finite remainder.

\subsubsection*{Structure of the poles}

The pole terms of the three-loop contribution to the self-energy can be predicted from the lower loops and, as expected, are of polylogarithmic type and rather compact. The first non-vanishing term in the expansion appears at order $\epsilon^{-2}$. Using the notation introduced in eq.~\eqref{eq:pi eps expansion}, the poles in terms of iterated integrals are given by
\begin{align}
    \Pi^{(3, -2)}(x) =& \frac{x^2-67 x+414}{36 (x-4) x}-\frac{(8 x-23)  }{r_1(x) r_2(x)(4-x) x}I(f_4;x) \, ,\\
    \Pi^{(3, -1)}(x) =&-\frac{(7 x+32) r_1(x) r_2(x) I(f_1,f_4;x)}{18 x^2}
    -\frac{\left(7 x^3-24 x^2-216 x+719\right)  I(f_3,f_4;x)}{9 r_1(x) r_2(x) (4-x) x}\nonumber\\
    &+\left(\frac{32}{9 x^2}+\frac{1}{9}\right) \Big[ I(f_4,f_1,f_4;x)
    +2 I(f_4,f_3,f_4;x)
    - I(f_1,f_4,f_4;x)  \nonumber\\
    & - 2 I(f_3,f_4,f_4;x) \Big]
    -\frac{\left(21 x^3-26 x^2-164 x+592\right) I(f_4,f_4;x)}{18 (x-4) x^2}\nonumber\\
    &-\frac{\left(2 x^3-130 x^2+1337 x-2388\right) I(f_4;x) }{24 r_1(x) r_2(x) (4-x) x}+\frac{361 x^2-8542 x+38760}{864 (x-4) x}\,.
\end{align}
As one would expect, the $\epsilon^{-2}$ contribution involves only the single kernel present in the one-loop finite remainder, and the $\epsilon^{-1}$ term involves only the three kernels present in the finite remainder at two loops (cf. eqs.~\eqref{eq:kernels 1 loop prop} and~\eqref{eq:kernels 2 loop prop}).

For completeness, we also give here the pole contributions at one and two loops. In both cases, there are only simple poles in $\epsilon$, and their coefficients read
\begin{align}
    \Pi^{(1,-1)}(x) = -\frac{1}{3 } \, , \qquad\qquad
    \Pi^{(2,-1)}(x) = -\frac{x+24}{8 x}+\frac{6 }{r_1(x) r_2(x) x}I(f_4;x)\,.
\end{align}

\subsection{Local solution close to $p^2=0$}

A special kinematical limit that is particularly important for physical purposes is $x\to 0$, which corresponds to studying the behavior of the 
photon self-energy when the photon goes on-shell. This limit
can be used to extract the photon wave function renormalization constant in QED to three loops in the on-shell scheme.
Here, we provide just the leading behavior for $x\to 0^+$, which is sufficient to obtain the wave function renormalization constant. In the ancillary files attached to this arXiv submission we include the expansion close to $x= 0$ up to order $100$, as well as expansions around other singular points to cover the whole kinematic space, as discussed in~\cref{sec:analytic cont}.

At $p^2=0$, the contributions up to three loops are simply
\begin{align}\label{eq:prop at zero}
    \Pi^{(1)}|_{p^2=0} &= -\frac{1}{3\epsilon}\,,\nonumber\\
    \Pi^{(2)}|_{p^2=0} &= \frac{3}{8 \epsilon }-\frac{13}{48}+\frac{35 \epsilon }{96} + \mathcal{O}(\epsilon^2)\,,\\
    \Pi^{(3)}|_{p^2=0} &= -\frac{1}{18 \epsilon ^2}-\frac{143}{216 \epsilon }+\left( \frac{281}{648} + \frac{37 }{96}\zeta (3) \right)+ \mathcal{O}(\epsilon)\,.\nonumber
\end{align}
Note that the one-loop result is exact in $\epsilon$, whereas we give the two-loop and three-loop results to the order required for the photon wave function renormalization.

\section{Renormalization}
\label{sec:ren}

We start with a few general remarks on the photon wave function renormalization, that can be found in a similar fashion in \cite{Grozin:2005yg}, for example. First, we re-define the photon field as
\begin{align}
    A^\mu = Z_3^{\frac{1}{2}}A^\mu_r\,.
\end{align}
From the photon self-energy, one can deduce the wave function renormalization constant $Z_3$ by considering the limit $p^2 \to 0$ or, equivalently, $x\to 0$.

By $D_{\gamma,\, r}^{\mu\nu} (p,m)$ we denote the propagator, including 1PI contributions, where the photon field has been renormalized while still using the bare charge and bare mass. This object is then related to the bare propagator in~\eqref{eq:photon prop} by
\begin{align}\label{eq:photon prop renorm}
    D_{\gamma,\, r}^{\mu\nu} (p,m) = Z_3^{-1} D_{\gamma,\, \text{bare}}^{\mu\nu} (p,m) = -\frac{i}{p^2} \left[\frac{Z_3^{-1}}{1-\Pi(p,m)}\left( g^{\mu\nu}-\frac{p^\mu p^\nu}{p^2}\right) + \xi \frac{p^\mu p^\nu}{p^2}\right] \, .
\end{align}
We impose the on-shell renormalization conditions, i.e., the pole at $p^2=0$ to have unit residue, which implies
\begin{align}\label{eq:renorm condition implicit}
    \frac{Z_3^{-1}}{1-\Pi(p,m)|_{p^2=0}} = 1 \, .
\end{align}
Expanding the inverse renormalization constant $Z_3^{-1}$ as
\begin{align}\label{eq:z3 expansion}
    Z_3^{-1} = 1 + \sum\limits_{l=1}^\infty \left(\frac{\alpha}{\pi} C(\epsilon)\right)^{l} Z_{3,\text{inv}}^{(l)} \, ,
\end{align}
and inserting~\eqref{eq:form factor expansion} and~\eqref{eq:z3 expansion} into~\eqref{eq:renorm condition implicit}, we find
\begin{equation}\label{eq:expand z3 in bare coupling}
     Z_{3,\text{inv}}^{(l)} = -\Pi_0^{(l)} \qquad \forall \; \ell ,
\end{equation}
where we introduced $\Pi _0^{(l)} =\Pi^{(l)}(p,m)|_{p^2=0}$ and used $C(\epsilon)$ as defined in~\eqref{eq:definition C}. These conditions relate the photon field renormalization constants to our results for the photon self-energy, evaluated at zero momentum. From the limiting behaviour for $x\to 0^+$ given in~\eqref{eq:prop at zero}, we find
\begin{align}
    Z_{3,\text{inv}}^{(1)} &= \frac{1}{3\epsilon} \, ,\nonumber\\
    Z_{3,\text{inv}}^{(2)} &= -\frac{3}{8\epsilon}+\frac{13}{48}-\frac{35 \epsilon }{96}+O(\epsilon^2 ) \, ,\\
    Z_{3,\text{inv}}^{(3)} &= \frac{1}{18\epsilon^2}+\frac{143}{216\epsilon} - \left(\frac{281}{648}+\frac{37 \zeta(3)}{96}\right)+O(\epsilon ) \, ,\nonumber
\end{align}
which is in agreement with~\cite {Broadhurst:1991fi} upon adjusting for the different overall normalization.
Here, we have presented the coefficients of $Z_3^{-1}$ to the order in $\epsilon$ required for renormalization to three loops. Note that since the photon self-energy to three loops is IR finite, we expect the complete UV renormalization to remove all of its poles in $\epsilon$.
Using our results, together with the electron mass renormalization constant found in~\cite{Melnikov:2000zc, Duhr:2024bzt} and the well-known fact that due to QED gauge invariance
\begin{align}\label{eq:relation bare and renorm coupling}
    \alpha = Z_3^{-1} \alpha_r\,,
\end{align}
 we confirm that after wave function, mass, and charge renormalization, the photon self-energy to three loops is indeed finite, which provides a cross-check of our calculation. As a second cross-check, expanding our fully renormalized self-energy close to $x=0$ and, again, accounting for a different overall normalization, we note that our result is consistent with the first three orders given in~\cite{Baikov:1995ui}.

\section{Analytic continuation}
\label{sec:analytic cont}

In this section, we describe the analytic continuation of the photon self-energy $\Pi^{(3)}(x)$ on the whole phase space, $x\in \mathbb R+i 0^+$. For this, we provide local generalized series expansions around the singular points of the differential equation $x_0\in\{0,1,4,16,\infty \}$. We follow the same strategy described in detail for the electron self-energy at three loops in reference~\cite{Duhr:2024bzt}. In particular, at each singular point, we solve the system of differential equations fixing the corresponding boundary conditions using a high-precision numerical evaluation obtained with \texttt{AMFlow}. To estimate the numerical error of our series expansions, we compare the agreement of two local series expansions inside their overlapping region of convergence. We stress that we perform the whole analytic continuation numerically and do not try to derive the boundary values analytically, except for the physically relevant point $x=0$, which has already been discussed in detail in the previous section.

Since the master integrals might develop logarithmic or algebraic branch cuts at the various singular points, to numerically evaluate the master integrals with \texttt{AMFlow}, we slightly perturb each singular point and evaluate the master integrals at $x=x_0\pm \delta$, for which we have chosen $\delta\sim 10^{-2}$. In this way, we obtain finite values for the master integrals close to each singular point. The differential equations with their series solution to high expansion order are then used to transport the boundary values back to $x=x_0$. In more detail, we expand the master integrals around the singularity $x_0$ as
\begin{equation}
    \vec{J}(x) = \sum_{n=0}^\infty \epsilon^n\, \vec{J}^n(x)
\end{equation}
with 
\begin{equation}
    \vec{J}^{(n)}(x) = \sum_{a\in I_a} \sum_{b\in I_b} \vec{C}_{a,b}^{(n)}\,  (x-x_0)^{a} \log^b{(x-x_0)}\,.
\label{eq:expJ}
\end{equation}
Notice that the number of allowed logarithms, i.e., the set $I_b$, is fixed by the differential equation at every order in $\epsilon$. In our case, the set $I_a$ runs over half-integers, and its length determines the precision of our series expansion, which can be arbitrarily increased. Explicitly evaluating the series at the value $x=x_0+10^{-2}$, we get the boundary values with a precision that is related to the expansion order. We can estimate their numerical precision by comparing the boundary values for different truncation orders.

While the procedure described above to obtain series expansion solutions from the differential equations does not strictly rely on a canonical basis, having one simplifies this construction substantially. Most importantly, this is the case because we expect our integrals to degenerate to simple integrals when special limits are considered in which the K3 geometry becomes singular. This form of the equations helps then to keep the complexity of the expansion under control and to easily obtain resummed results close to regular singular points, if necessary. A delicate point in this construction is that, as we already discussed, our canonical basis is strictly speaking only defined locally close to $x_0=0$. As a next step, we would like to have a procedure to extend the definition of our canonical basis to any other point so that we can use it to derive series expansion solutions valid on the whole kinematic space. In doing that, we would like to use a rotation which, at every point, looks formally identical to the one derived at $x=0$, see Appendix~\ref{app:canbasis}. We specify what we mean by this in the following paragraphs.

 Clearly, as long as we have to deal with rational functions, we can simply substitute $x$ by the new local variable, i.e., we replace $x\rightarrow x-x_0$ (or $x\rightarrow 1/x$). We do the same for the appearing square roots but also redefine them to be positive for (small) positive values of the new local variable. More precisely, we use the following square roots
\begin{alignat}{3}
    &\left\{ \sqrt{x},\; \sqrt{4-x}, \; \sqrt{16-x} \right\}\, ,\quad &&\mbox{at}\;  x=0  && \nonumber \\
    &\left\{ \sqrt{1+u},\; \sqrt{3-u},\; \sqrt{15-u} \right\}\, ,  \quad &&\mbox{at}\; x=1\; &&\mbox{with}\quad u=x-1 \,, \nonumber \\
    &\left\{ \sqrt{4+v},\; \sqrt{v},\; \sqrt{12-v} \right\}\, ,   \quad &&\mbox{at}\; x=4 && \mbox{with} \quad v= x-4\,,  \nonumber  \\
    &\left\{ \sqrt{16+w},\; \sqrt{12+w},\; \sqrt{w} \right\}\, ,\quad 
    && \mbox{at}\; x=16 && \mbox{with} \quad w = x-16\,,  \nonumber \\
    &\left\{ \sqrt{z}, \; \sqrt{1-4z},\; \sqrt{1-16z} \right\}\, ,\quad && \mbox{at}\; x=\infty \quad && \mbox{with} \quad z = 1/x  \,,
\end{alignat}
close to each regular or regular singular point defined in section~\ref{sec:geo}.

For the transcendental functions, particularly for $\varpi_0$, we have to be more careful. As indicated in section~\ref{sec:geo}, there are two important points to consider. First,  there is no canonical choice of $\varpi_0$ at a non-MUM point. Second, even if one only considers MUM points (say $\varpi_0^{[0]}$ at $x=0$), it is well known that the analytic continuation of $\varpi_0^{[0]}$ to another MUM point will, in general, involve a mixing of all three periods defined at that point. In other words, the analytic continuation of the holomorphic period defined at $x=0$ to a different point $x=x_0$ will, in general, also involve logarithms $\sim \log^n{(x-x_0)}$, with $n \leq 2$.

We propose here a general set of conditions that allows us to define a consistent choice of $\varpi_0^{[x_0]}$ at any point $x_0$, starting from the original choice of $\varpi_0^{[0]}$ that we made at the MUM point $x=0$. We explicitly demand the following requirements:
\begin{enumerate}
    \item First of all, the choice of $\varpi_0^{[x_0]}$ should guarantee that there are at most single poles in the differential equation at $x=x_0$.
    
    \item Second, $\varpi_0^{[x_0]}$ should be holomorphic (a simple power series) close to $x=x_0$, such that we do not introduce any mixing of transcendental weights. 
    
    \item For the last and most important property, let us recall that in the construction of the canonical basis, we never used the explicit form of $\varpi_0$ but only the relations it satisfies. In particular, we used Griffiths transversality to remove its second derivative, see eq.~\eqref{eq:griffiths}. If we want to be able to use the very same form of the canonical differential equations at $x=x_0$, we then need to ensure that our choice of $\varpi_0^{[x_0]}$ satisfies the same relation after replacing $x\rightarrow x-x_0$ (or $x\rightarrow 1/x$).
\end{enumerate}

These three constraints usually do not fix $\varpi_0^{[x_0]}$ uniquely. Nevertheless, we expect that any choice of $\varpi_0^{[x_0]}$ satisfying the three mentioned conditions is equally valid as an ingredient for extending the canonical basis from $x=0$ to $x=x_0$. Having appropriately chosen $\varpi_0^{[x_0]}$, we can use it to compute the $G_i$ functions following eq.~\eqref{eq:def G functions} with the corresponding replacements for the rational and algebraic functions. In this way, we can derive a canonical basis for the differential equations, valid locally close to any point (regular or singular). In section~\ref{sec:geo}, we have already chosen $\varpi_0^{[x_0]}$ such that it satisfies the above-mentioned conditions. One can easily check that these choices indeed satisfy the Griffiths relation \eqref{eq:griffiths} in the corresponding local variables. For example, for $v=x-4$, we find
\begin{equation}
    {\varpi_0^{[4]}}''(v) = -\frac{4-v}{2 (12-v) (4+v) v}\varpi_0^{[4]}(v)-\frac{2 \left(12+7 v-v^2\right)}{(12-v) (4+v) v}{\varpi_0^{[4]}}'(v)+\frac12\frac{{\varpi_0^{[4]}}'(v)^2}{\varpi_0(v)}  \, ,
\end{equation}
which is true for the choice of $\varpi_0^{[4]}$ given in eq.~\eqref{eq:varpiv}. From this, we get the following expansions of the $G_i$ functions:
\begin{equation}
\begin{aligned}
    G_1^{[4]}(v) &= -\frac{3}{2 v}-\frac{v}{48}-\frac{v^2}{180}-\frac{11 v^3}{69120}-\frac{31 v^4}{414720} + \mathcal O(v^5) \, , \\
    G_2^{[4]}(v) &= -\frac{\sqrt{3}}{\sqrt{v}}-\frac{\sqrt{3} \sqrt{v}}{8}+\frac{17 v^{3/2}}{384 \sqrt{3}}-\frac{217 v^{5/2}}{76800 \sqrt{3}}+\frac{2423 v^{7/2}}{3440640 \sqrt{3}}-\frac{11209 v^{9/2}}{106168320 \sqrt{3}}   + \mathcal O(v^{11/2}) \, , \\
    G_3^{[4]}(v) &= -\frac{6}{\sqrt{v}}-\frac{\sqrt{v}}{4}+\frac{5 v^{3/2}}{192}-\frac{437 v^{5/2}}{115200}+\frac{491 v^{7/2}}{737280}-\frac{2233 v^{9/2}}{17694720}   + \mathcal O(v^{11/2}) \, .
\end{aligned}
\end{equation}

\begin{figure}[h]
\begin{subfigure}{.5\textwidth}
    \centering
    \includegraphics[width=0.9\linewidth]{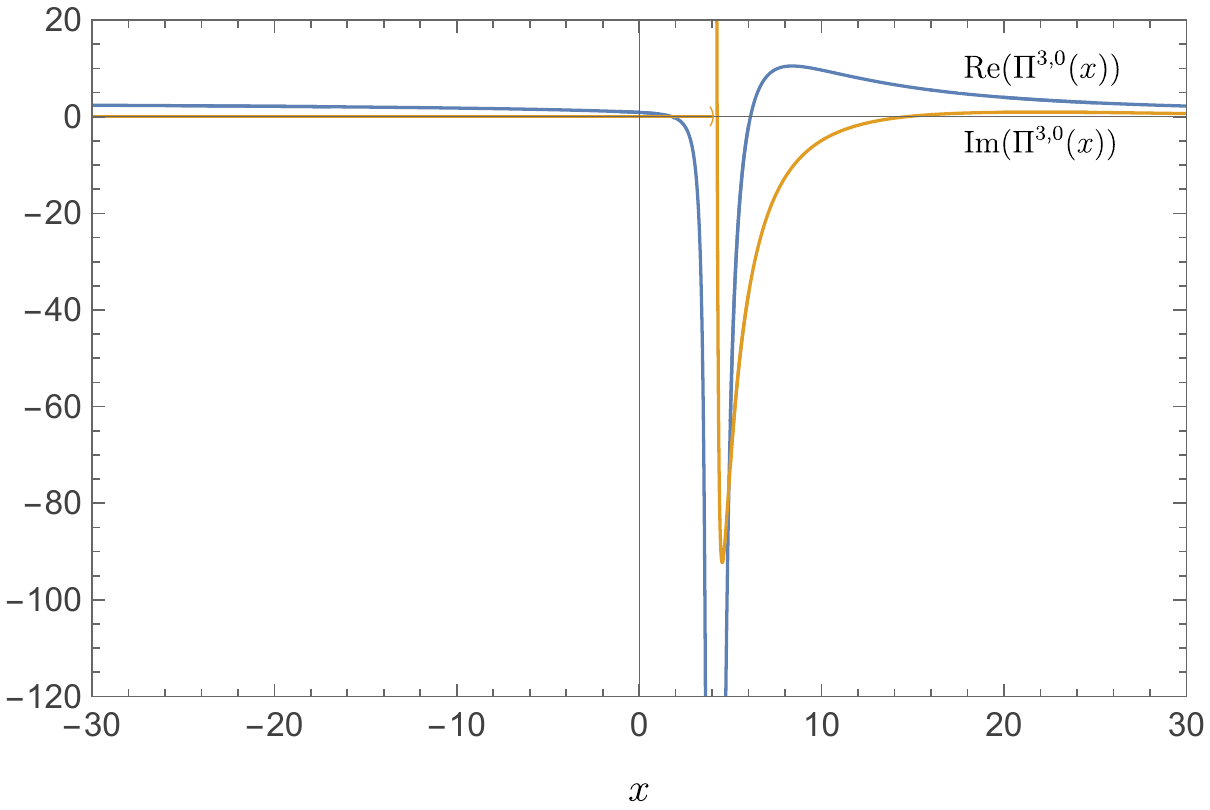}
\end{subfigure}    
\begin{subfigure}{.5\textwidth}
    \centering
    \includegraphics[width=0.9\linewidth]{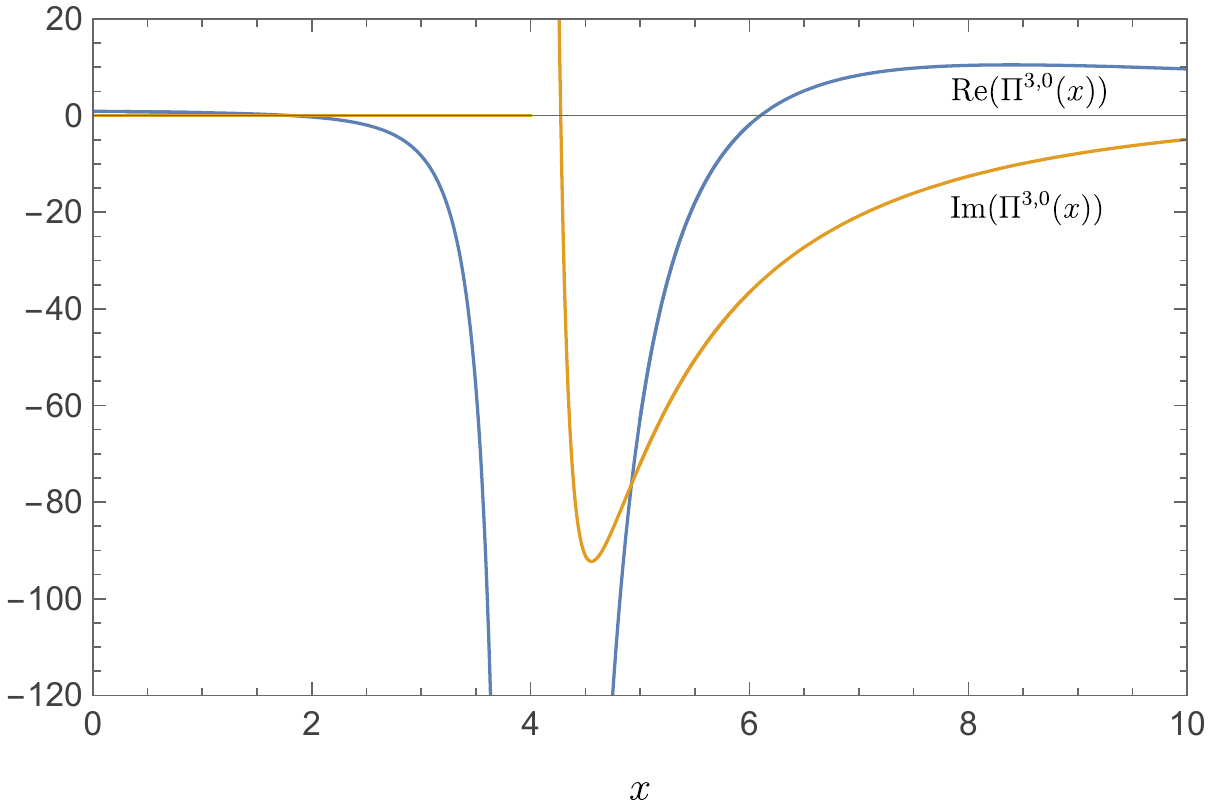}
\end{subfigure}    
\caption{Real and imaginary part of $\Pi^{3,0}(x)$ across the whole kinematic space (left panel) and a zoom of the region close to the threshold $x=4$ (right panel).}\label{fig:plot}
\end{figure}

Importantly, the local series expansions around the singular points $x_0\in\{0,1,4,16,\infty \}$ are not sufficient to cover also the region $x\in(-16,-1)$. To remedy this issue, we follow the same strategy as in~\cite{Duhr:2024bzt} and use a Möbius transformation
\begin{equation}
    \eta = \frac x{2-x}\, \qquad\text{with inverse}\qquad x = \frac{2\eta}{1+\eta} \, .
\end{equation}
In this way, we map $\{\infty,0,1\}$ to $\{-1,0,1\}$ and we can use the series expansion around $x=0$ expressed through the Möbius variable $\eta$ to evaluate the photon self-energy also on the interval $x\in(-16,-1)$. This allows us to cover the full kinematic space with generalized power series. To visualize the analytic properties of the photon self-energy, we plot the $\mathcal{O}(\epsilon^0)$ contribution $\Pi^{3,0}(x)$ across the whole kinematic space as shown in Fig.~\ref{fig:plot}. 

Let us make some comments on the results. It turns out that the series expansions of the master integrals and of the self-energy at $x=0$ have a larger radius of convergence than expected from the differential equation. In particular, the singularity of the differential equations at $x=1$ seems not to restrict the radius of convergence. This goes along with similar observations made in the context of simpler problems~\cite{Pozzorini:2005ff} and can be traced back to the fact that the physical solutions for all master integrals are regular at $x=1$. The self-energy develops a square-root branch cut from the threshold $x=4$ to $x=\infty$. Before the threshold, the self-energy is purely real. Interestingly, at the singularity $x=16$, the self-energy is also regular, whereas the individual master integrals develop a further square root cut which cancels in the final physical result.

Before closing our discussion, we want to comment on the use of so-called Bernoulli-like variables~\cite{tHooft:1978jhc} to obtain accelerated series expansions for our result. These variables are typically used to improve the convergence of classical and multiple polylogarithms~\cite{Gehrmann:2001jv, Vollinga:2004sn, Buehler:2011ev}, but their utility was also demonstrated in elliptic cases, see~\cite{Pozzorini:2005ff} and, more recently in reference~\cite{Duhr:2024bzt}. Let us consider, in particular, the series solution centered at $x=0$. Since we can ignore the closest singularity at $x=1$, the next relevant singularity that would restrict its radius of convergence is $x=4$. By transforming to the Bernoulli-like variable 
\begin{equation}
    z_B = -\log\left( 1-\frac x4 \right)\, \quad\text{whose inverse reads} \quad x =4(1-e^{-z_B}) \, ,
\end{equation}
this singularity is pushed to infinity. Similar Bernoulli-like variables can also be defined at the other relevant singular points, allowing us to cover the whole kinematic space with higher numerical precision but still with the same number of terms. 
 
 We stress here that, due to the fact that our differential equations have, in general, more than three singular points, it is not obvious that this transformation should improve the convergence of the series. 
 In fact, whenever we are dealing with a problem characterized by only three singular points (which, for definiteness, can always be fixed to be $\{0,1,\infty\}$), it is always possible to redefine the series close to $x=0$ with a Bernoulli-like variable that pushes $x=1$ to infinity, extending in this way the expected radius of convergence of the series. 
 If a fourth singular point exists, under a standard Bernoulli transformation, this will typically be moved to some location in the complex plane at a finite distance from $x=0$. As a result, the radius of convergence of the series expansion will still be limited. Despite this, we still observe a substantial improvement in the convergence of the series. It will be interesting to investigate in the future the general applicability of these changes of variables and their possible generalization to cases with multiple singularities, such as the ones considered here.
 
 In the ancillary file to the arXiv submission, we include the expansions in the Bernoulli-like variable up to 100 orders, which gives a precision better than 13 digits across the whole parameter space. Compared to the electron self-energy, this convergence is substantially slower. We can trace this behaviour back to the underlying K3 geometry, for whose periods the Bernoulli-like variables work less well than for elliptic and polylogarithmic counterparts. We leave it to future investigations to study this behaviour in greater generality for K3 and Calabi-Yau geometries.


\section{Conclusions}
\label{sec:conc}

In this paper, we have computed the three-loop photon self-energy in QED and analyzed its properties as a function of the momentum squared. 
We have performed our study analytically, but we also provided a thorough discussion of the numerical evaluation of our results across the whole kinematic space. Following the traditional approach, we have started with the relevant Feynman diagrams and have used IBP relations to derive the necessary scalar photon self-energy in terms of master integrals. Through an analysis of the corresponding maximal cuts, we have classified the master integrals as either of polylogarithmic or K3 type. Here, the K3 geometry was traced back to the three-loop equal-mass banana graph. We have computed the appearing master integrals using the method of differential equations supplemented with an $\epsilon$-factorized basis. This was accomplished by applying recent techniques for deriving $\epsilon$-factorzied bases. Moreover, we have argued that the resulting basis also features all necessary conditions to be the right generalization of a canonical basis beyond the polylogarithmic case. 
With this basis, we were able to identify a set of linearly independent integration kernels, which we then used to write down our master integrals in terms of independent (Chen) iterated integrals. The boundary values for these integrals have been calculated using high-precision numerical evaluations obtained with \texttt{AMFlow} close to regular singular points, supplemented by an integer relation algorithm to map them to combinations of numbers of polylogarithmic origin. Some of the master integrals were expressed as iterated integrals over kernels containing K3 periods, which gives rise to a new class of mathematical functions. 
Additionally, having chosen a basis of independent iterated integrals was crucial to prove the cancellation of a large number of differential forms in the final $\mathcal{O}(\epsilon^0)$ result for the self-energy and also to analytically perform UV renormalization. As expected, all poles in $\epsilon$ were of polylogarithmic origin and could be obtained by the one- and two-loop orders, respectively. In this context, we have re-derived the QED photon-field renormalization constants in QED up to three loops. Finally, we have covered the whole kinematic space with generalized power series expansions located at the singularities of the differential equation. Bernoulli-like variables have been used to accelerate the convergence of these series. Their evaluation is nearly instantaneous and allows us to obtain high-precision numerical values for the self-energy for all values of the momentum squared. 

 The calculation of the photon self-energy at three loops serves as an interesting example of a well-defined quantity in quantum field theory which features functions beyond the polylogarithmic case. We have demonstrated that our analytic and numerical approach is powerful enough to tackle problems of K3 type. The whole analysis in this paper, along with a related work addressing the electron self-energy at three loops done by some of the authors, concludes the analytic discussion of self-energies in QED up to three loops. This progress obviously suggests the extension of this research to the next loop level, where we expect an even richer mathematical structure related to higher-dimensional Calabi-Yau geometries. We leave this for future work. Additionally, it will be worthwhile to elaborate more rigorously on the use of Bernoulli-like variables for general Fuchsian differential equations. In this context, one can consider the obvious generalizations either in the directions of arbitrary Calabi-Yau differential equations or in more general situations where the differential equations exhibit multiple singularities (which could be of Calabi-Yau, polylogarithmic or other types). Both avenues have the potential to enhance the reach and applicability of series expansions for the numerical evaluation of Feynman integrals.

\acknowledgments
We thank David Broadhurst, Federico Buccioni, Albrecht Klemm and Nikolaos Syrrakos for insightful discussions. We are also grateful to Claude Duhr, Federico Gasparotto, Sara Maggio and Stefan Weinzierl for collaboration on related projects and we are especially indebted to Fabian Wagner for his comments, discussions, and collaboration during the initial phases of this project. 
We are grateful to the Munich Institute for Astro-, Particle and BioPhysics (MIAPbP), funded by the Deutsche Forschungsgemeinschaft (DFG, German Research Foundation) under Germany´s Excellence Strategy – EXC-2094 – 390783311, where part of these results have been obtained.
LT and CN were supported in part by the Excellence Cluster ORIGINS 
funded by the Deutsche Forschungsgemeinschaft (DFG, German Research Foundation) under Germany’s 
Excellence Strategy – EXC-2094-390783311 and in part by the European Research Council (ERC) under the European Union’s research and innovation program grant agreements 949279 (ERC Starting Grant HighPHun). 
FF was supported by the research unit FOR 5582 funded by the Deutsche Forschungsgemeinschaft (DFG, German Research
Foundation) – Projektnummer 508889767.

\appendix

\section{Canonical basis}
\label{app:canbasis}

In this appendix, we present our choice of canonical master integrals. We set $m=1$ for simplicity and use the dimensionless variable $x=p^2/m^2$. Furthermore, we work in the region $0<x<4$. The rotation involves the three square roots 
\begin{equation}
    r_1(x)=\sqrt{x}\,,\quad r_2(x)=\sqrt{4-x}\,,\quad r_3(x)=\sqrt{16-x} \, ,
\end{equation}
the K3 period $\varpi_0(x)$ discussed in section~\ref{sec:geo} and the three new transcendental functions
\begin{align}
G_1(x) &\coloneqq -\int\limits_0^x \mathrm d u \frac{(u-8) (u+8)^3 \varpi_0(u)^2}{32 (u-16)^2 (u-4)^2}\,,\nonumber\\
    G_2(x) &\coloneqq \int\limits_0^x \mathrm d u \frac{8 G_1(u) r_3(u) r_2(u)}{(4-u) (16-u) u \varpi_0(u)}\, , \nonumber\\
    G_3(x) &\coloneqq \int\limits_0^x \mathrm d u \frac{(u+2) r_1(u) r_2(u) \varpi_0(u)}{(u-4)^2 u} \, .
\end{align}
To relate integrals defined in four dimensions to two-dimensional ones, we use the dimensional shift operator $\dim$. More specifically, by ${\bf D}^- I$, we refer to the integral $I$ evaluated in two dimensions, expressed in terms of the set of four-dimensional basis integrals. In this way, we can search for good candidates also in two dimensions, while consistently using four-dimensional integrals for the whole set of basis integrals.

Before listing our complete basis, we discuss how to $\epsilon$-factorize sector 202 of family $B$, which is given by the three-loop banana graph, following the procedure described in~\cite{Gorges:2023zgv}. We focus on this sector since it is the only one whose homogeneous solutions introduce a new geometry, namely, the K3 geometry in this case.

To obtain a set of canonical integrals for the banana sector, we concatenate several rotations,
\begin{align}
    (J_{22},\, J_{23},\, J_{24})^T    =   T_G\, T_c\, W_\text{ss}^{-1}\, T_d \; ({\bf D}^- I^B_{0 1 0 1 0 0 1 1 0},\, {\bf D}^- I^B_{0 2 0 1 0 0 1 1 0},\, {\bf D}^- I^B_{0 3 0 1 0 0 1 1 0})^T,
\end{align}
where
\begin{equation}
\begin{aligned}
    T_d    &=  \left(  \begin{array}{ccc}
                                        1 & 0 & 0 \\
                                        -\frac{1+3 \epsilon }{x} & -\frac{4}{x} & 0 \\
                                        \frac{(1-\epsilon ) (1+3 \epsilon )}{x^2} & \frac{4+x-(4-x) \epsilon }{x^2} & \frac{2}{x} \\
                        \end{array} \right) \,,\\
    W_\text{ss}   &=  \left(  \begin{array}{ccc}
                                        \varpi_0(x) & 0 & 0 \\
                                        \varpi_0(x)' & R(x) & 0 \\
                                        \varpi_0(x)'' & R(x)\frac{\varpi_0'(x)}{\varpi_0(x)} + R'(x) & \frac{64}{(16-x) (4-x) x^2 \varpi_0(x)} \\
                        \end{array} \right) \,, \\
    T_c    &=  \left(  \begin{array}{ccc}
                                        \eps^2 & 0 & 0 \\
                                        \frac{(10-x) x r_1r_2\epsilon^2 }{2 (16-x) (4-x)}\varpi_0 & \eps & 0 \\
                                        (R_1\eps+R_2\eps^2)\varpi_0^2-\frac{1}{32} (10-x) x^2\varpi_0\varpi_0' & \frac{(10-x) x r_1r_2\epsilon }{2 (16-x) (4-x)}\varpi_0 & 1 \\
                        \end{array} \right) \,, \\ 
    T_G    &=  \left(  \begin{array}{ccc}
                                        1 & 0 & 0 \\
                                        G_2(x) & 1 & 0 \\
                                        \frac{G_1(x)}{\eps}-\frac{G_2(x)^2}2 & -G_2(x) & 1 \\
                        \end{array} \right) \,,
\end{aligned}
\end{equation}
and we abbreviated the rational functions
\begin{equation}
\begin{aligned}
    R(x)    &=  \frac{8r_1(x)r_2(x)}{(16-x) (4-x) x}    \,, \\
    R_1(x)  &=  -\frac{\left(640-228 x+30 x^2-x^3\right) x}{32 (16-x) (4-x)}    \,, \\
    R_2(x)  &=  \frac{4096+512 x-800 x^2+168 x^3-7 x^4}{128 (16-x) (4-x)}   \, .
\end{aligned}
\end{equation}

The first rotation $T_d$ is trivial, as it only amounts to bringing our system into its derivative basis. Following~\cite{Gorges:2023zgv}, the main step is then to multiply by the inverse semi-simple part of the Wronskian matrix of the banana integrals, $W_\text{ss}^{-1}$. This can be understood as a generalization of removing leading singularities, and it allows to separate master integrals with different transcendental weights close to $x=0$. Next, we rotate by $T_c$ as a cleanup step to remove contributions of $\varpi_0'$ and $\varpi_0''$ in the differential equations. Notice that this step also involves a rescaling by different powers of $\epsilon$, which is required to compensate for the different transcendental weights of the three masters after the semi-simple rotation. 
Finally, one further rotation $T_G$ is required to reach the desired $\epsilon$-factorization. Here, it is necessary to introduce the functions $G_1$ and $G_2$ defined in~\eqref{eq:def G functions}. We stress here that an equivalent basis can be obtained following the Ansatz procedure proposed in~\cite{Pogel:2022vat}.

Having discussed the banana sector, we now provide the rest of our canonical basis. To obtain a unique ordering among the basis integrals, we order them first by their family and then by increasing number of different propagators in the denominator. Among integrals where this number coincides, we order by increasing sector ID, defined in \cref{eq:sec id}. Then, the full canonical basis, excluding the banana sector discussed above, reads
\begin{alignat}{2}
&\mbox{FA Sector 7:}    \quad 
    & J_{1} &=  
    - \eps^3 \; {\bf D}^- I^A_{1 1 1 0 0 0 0 0 0}   \, ,
\nonumber \\
&\mbox{FA Sector 14:}   \quad
    & J_{2} &= 
    - \eps^3\, r_1r_2 \; {\bf D}^- I^A_{1 1 1 1 0 0 0 0 0}  \, ,
\nonumber \\
&\mbox{FA Sector 78:}    \quad 
    & J_{3} &=  
    - \eps^3\, r_1r_2 \; {\bf D}^- I^A_{0 1 1 1 0 0 1 0 0}  \, ,
\nonumber \\
&    
    & J_{4} &= 
    - \eps^3 \; {\bf D}^- I^A_{-1 1 1 1 0 0 1 0 0}  \, ,
\nonumber \\
&\mbox{FA Sector 198:}    \quad 
    & J_{5} &=  
    -(1+4 \epsilon ) \epsilon ^2 \; {\bf D}^- I^A_{0 1 1 0 0 0 1 1 0}  \, ,
\nonumber \\
&\mbox{FA Sector 212:}    \quad 
    & J_{6} &=  
    - \eps^3\, r_1r_2 \; {\bf D}^- I^A_{0 0 1 0 1 0 1 1 0}  \, ,
\nonumber \\
&    
    & J_{7} &= 
    \eps^3 \; ({\bf D}^- I^A_{-1 0 1 0 1 0 1 1 0} - {\bf D}^- I^A_{0 0 1 0 1 0 1 1 0})  \, ,
\nonumber \\
&    
    & J_{8} &= 
    - \eps^3 \; {\bf D}^- I^A_{0 -1 1 0 1 0 1 1 0}  \, ,
\nonumber \\
&\mbox{FA Sector 31:}    \quad 
    & J_{9} &=  
    - \eps^3\, x(4-x) \; {\bf D}^- I^A_{1 1 1 1 1 0 0 0 0}  \, ,
\nonumber \\
&\mbox{FA Sector 110:}    \quad 
    & J_{10} &=  
    - \eps^3\, x(4-x) \; {\bf D}^- I^A_{0 1 1 1 0 1 1 0 0}  \, ,
\nonumber \\
&    
    & J_{11} &= 
    - \eps^3\, r_1r_2 \; {\bf D}^- I^A_{-1 1 1 1 0 1 1 0 0}  \, ,
\nonumber \\
&\mbox{FA Sector 206:}    \quad 
    & J_{12} &=  
    - \eps^3\, r_1r_2 \; {\bf D}^- I^A_{-1 1 1 1 0 0 1 1 0}  \, ,
\nonumber \\
&    
    & J_{13} &= 
    - \eps^3 \; {\bf D}^- I^A_{-2 1 1 1 0 0 1 1 0}  \, ,
\nonumber \\
&    
    & J_{14} &= 
    - \eps^3\, (1-x) \; ({\bf D}^- I^A_{0 1 1 1 0 0 1 1 -1} -2 \; {\bf D}^- I^A_{0 1 1 1 0 0 1 0 0})   \, ,
\nonumber \\
&\mbox{FA Sector 214:}    \quad 
    & J_{15} &=  
    - \eps^3\, r_1r_2 \; {\bf D}^- I^A_{0 1 1 0 1 0 1 1 -1}  \, ,
\nonumber \\
&\mbox{FA Sector 63:}    \quad 
    & J_{16} &=  
    - \eps^3\, x(4-x)r_1r_2 \; {\bf D}^- I^A_{1 1 1 1 1 1 0 0 0}  \, ,
\nonumber \\
&\mbox{FA Sector 222:}    \quad 
    & J_{17} &=  
    - \eps^4(1-2\eps)\, x \; I^A_{0 1 2 1 1 0 1 1 0}  \, ,
\nonumber \\
&\mbox{FA Sector 246:}    \quad 
    & J_{18} &=  
    - \eps^4(1-2\eps)\, x \; I^A_{0 1 1 0 1 1 2 1 0}  \, ,
\nonumber \\
&\mbox{FA Sector 246:}    \quad 
    & J_{19} &=  
    - \eps^5(1-2\eps)\, x \; I^A_{1 1 1 0 1 1 1 1 0}  \, ,
\nonumber \\
&    
    & J_{20} &= 
    - \eps^5(1-2\eps)\, xr_1r_2 \; I^A_{1 1 1 1 1 1 1 1 0}   \, ,
\nonumber \\
&\mbox{FB Sector 195:}    \quad 
    & J_{21} &=  
    - \eps^3 \; {\bf D}^- I^B_{1 1 0 0 0 0 1 1 0}  \, ,
\nonumber \\
%
%
%
%
%
%
%
%
%
%
%
%
&\mbox{FB Sector 203:}    \quad 
    & J_{25} &=  
    - \frac{\eps^3}{4} \, \left[r_1r_2 \; {\bf D}^- I^B_{1 1 0 1 0 0 1 1 0} -\left(\frac{G_3}{2\varpi_0} - \frac{3r_1r_2}{2(4-x)}\right) \; {\bf D}^- I^B_{0 1 0 1 0 0 1 1 0}\right]  \, ,
\nonumber \\
&\mbox{FB Sector 220:}    \quad 
    & J_{26} &=  
    - \eps^3 \, x \; (2 \; {\bf D}^- I^B_{-1 0 1 1 1 0 1 1 0} + (4-x) \; {\bf D}^- I^B_{0 0 1 1 1 0 1 1 0} - \; {\bf D}^- I^B_{1 1 0 0 0 0 1 1 0}   \nonumber   \\
&   &   &   \quad 
    -2 \; {\bf D}^- I^A_{0 1 1 1 0 0 1 0 0})  \, ,
\nonumber \\
&    
    & J_{27} &= 
    - \eps^3 \; ({\bf D}^- I^B_{-1 1 1 1 0 0 1 0 0} -\; {\bf D}^- I^A_{0 1 1 1 0 0 1 0 0})   \, ,
\nonumber \\
&\mbox{FB Sector 219:}    \quad 
    & J_{28} &=  
    - \eps^4(1-2\eps) \, x \; I^B_{1 1 0 1 1 0 2 1 0}  \, ,
\nonumber \\
&    
    & J_{29} &= 
    - \eps^3 \; \left[(1-2\eps) \, x \; I^B_{2 1 0 1 1 0 2 1 0} -\frac{xr_1r_2}6\; {\bf D}^- I^B_{1 1 0 1 0 0 1 1 0}    \right.   \nonumber   \\
&   &   &   \quad \left.
    + \left(\frac{2G_3}{3\varpi_0} - \frac{(8+x)r_1r_2}{6(4-x)}\right) \; {\bf D}^- I^B_{0 1 0 1 0 0 1 1 0} +\frac{r_1r_2}6 \; {\bf D}^- I^B_{1 1 0 0 0 0 1 1 0}\right]   \, ,
\nonumber \\
&\mbox{FB Sector 221:}    \quad 
    & J_{30} &=  
    - \eps^4(1-2\eps) \, x \; I^B_{1 0 1 1 2 0 1 1 0}  \, ,
\nonumber \\
&    
    & J_{31} &= 
    \frac{r_1r_2}{s} \left(\frac{x}{\eps}\frac{\partial}{\partial x} - 1\right) \; J_{30} -\frac13\left[G_3 - \frac{(16-x)r_1r_2\varpi_0}{4(4-x)}\right] \; J_{22}   \, ,
\nonumber \\
&\mbox{FB Sector 462:}    \quad 
    & J_{32} &=  
    - \eps^4(1-2\eps) \, x \; I^B_{0 1 1 1 0 0 2 1 1}  \, ,
\nonumber \\
&    
    & J_{33} &= 
    \frac{r_1r_2}{s} \left(\frac{x}{\eps}\frac{\partial}{\partial x} - 1\right) \; J_{32} -\frac23\left[G_3 - \frac{(16-x)r_1r_2\varpi_0}{4(4-x)}\right] \; J_{22}   \nonumber   \\
&   &   &   \quad 
    +\frac{r_1r_2}{2(1-s)} \; J_{14}   \, ,
\nonumber \\
&    
    & J_{34} &= 
    - \eps^3(1-2\eps) \, x \; I^B_{0 1 1 1 0 0 3 1 1} +\frac{r_1r_2}{4(4-s)} \; J_{33}   \nonumber   \\
&   &   &   \quad 
    +\left[\frac{r_1r_2G_3}{6(4-x)} - \frac{(16-x)(4+x)\varpi_0}{48(4-x)}\right] \; J_{22} -\frac{1}{4(1-s)} \; J_{14}   \, ,
\nonumber \\
&\mbox{FB Sector 479:}  \quad
    & J_{35} &=
    - (1-2\eps)\eps^5 x\, r_1r_2 \; I^B_{1 1 1 1 1 0 1 1 1} \, ,
\nonumber \\
&    
    & J_{36} &= 
    - (1-2\eps)\eps^5( x \; I^B_{1 1 1 1 1 -1 1 1 1} -x^2 \; I^B_{1 1 1 1 1 0 1 1 1} )  \, .
\end{alignat}

\bibliographystyle{JHEP} 
\bibliography{biblio} 

\end{document}